\documentstyle[11pt]{article}
\begin{document}
\begin{titlepage}
\begin{flushright}
SUSSEX-AST 93/7-3\\
gr-qc/9307036\\
(July 1993)\\
\end{flushright}
\begin{center}
\Large
{\bf Can the Gravitational Wave Background from Inflation be
Detected Locally?}\\
\vspace{.3in}
\normalsize
\large{Andrew R. Liddle} \\
\normalsize
\vspace{.6 cm}
{\em Astronomy Centre, \\ School of Mathematical and Physical Sciences,\\
University of Sussex, \\ Brighton BN1 9QH, U.~K.}\\
\vspace{.6 cm}
\end{center}
\baselineskip=24pt
\begin{abstract}
\noindent
The Cosmic Background Explorer (COBE) detection of microwave background
aniso\-tro\-pies may contain a component due to gravitational waves generated
by inflation. It is shown that the gravitational waves from inflation might be
seen using `beam-in-space' detectors, but not the Laser Inter\-fer\-o\-meter
Gravity Wave Observatory (LIGO). The central conclusion, dependent only on
weak assumptions regarding the physics of inflation, is a surprising one. The
larger the component of the COBE signal due to gravitational waves, the {\em
smaller} the expected local gravitational wave signal.
\end{abstract}

\begin{center}
\vspace{1cm}
PACS numbers~~~98.80.Cq, 04.30.+x, 04.80.+z\\
\end{center}

\end{titlepage}

One of the many interesting possibilities highlighted by the observation of
cosmic microwave background aniso\-tro\-pies by the COBE satellite \cite{COBE}
is that the instrument may be directly seeing the influence of extremely long
wavelength gravitational waves \cite{TENSORS}. In general, the anisotropy seen
can arise as a combination of two contributions, the first being gravitational
waves and the second being genuine density perturbations in the matter
distribution at the time that the microwave background radiation was released.
A given cosmological theory predicts the relative significance of these in
influencing the microwave background.

One of the favoured models for the origin of inhomogeneities is the
inflationary paradigm \cite{INFL}, whereby the universe experienced a period
of accelerated expansion in the distant past. This is the only means by which
adiabatic density perturbations may be generated (rather than simply assumed
to be pre-existing), and it is fair to say that detailed analysis \cite{LL2}
of the consequences of the COBE observation appear to favour the perturbations
being adiabatic. Inflationary models also generate a stochastic spectrum of
gravitational waves modes, both spectra extending typically from wavelengths
of order centimeters (though the density perturbations on this sort of scale
do not survive long) all the way up to sizes much greater than the present
observable universe. The COBE measurement itself does not allow one to
distinguish between the two contributions to the anisotropy. However, the
scaling of the anisotropies with the angular resolution of observation is
different for the two types of contribution. Two or more experiments operating
on different angular scales (or a single one with a wide range of angular
resolutions) can therefore attempt to resolve the two contributions, and it
has been shown \cite{CBDES} that although this is not possible yet, it may
become so with improved data in the near future.

Reproducing the COBE measurement provides for the first time an accurate
normalization of the spectra for given inflationary models \cite{EPSINF}. For
a specific model, one therefore knows the entire gravitational wave spectrum,
and so can address the question of whether or not this stochastic background
might be detectable in proposed gravitational wave detectors. It might as well
be said straight away that the constraints have tightened to such an extent
that the case for the LIGO systems \cite{LIGO} currently under construction
seeing this background is already hopeless. The discussion here shall
therefore be directed primarily at the proposed `beam-in-space' experiments
\cite{BIS}. The sensitivity of these in terms of dimensionless gravitational
wave amplitude is likely if anything to be less than LIGO, but they probe
considerably longer wavelengths where the anticipated signal is larger, and
consequently one should be much more optimistic as to their utility in
searching for the inflationary background.

The key conclusion of this paper is a surprising one --- if gravitational
waves are identified as a component of the COBE signal, then it becomes harder
to detect them in our own solar system. Physically, this arises because
gravitational waves can only be significant if the inflationary expansion is
far from the de Sitter limit, implying a rapidly decreasing Hubble parameter.
This conclusion, though not previously stated in as explicit a manner, is
already apparent in studies which have restricted themselves to looking at
particular classes of inflationary models such as power-law inflation
\cite{SS}. Alternative studies made recently have been less clear on this
point, for example \cite{KW} by assuming that the entire COBE result is due to
gravitational waves and then not following through the implication that this
will necessitate a rapidly decreasing Hubble parameter. This {\em Letter} aims
to improve on this in three ways. Firstly, the discussion shall be kept in
general terms, without restriction to specific inflationary models or classes
of models. Secondly, the models shall be normalized to COBE accounting for
both contributions to the signal. Thirdly, a more exact treatment, based on
the Hamilton--Jacobi formalism, is used to consider the evolution of the
Hubble parameter from the time at which the modes relevant for the COBE
normalization are generated up to the corresponding time for modes detectable
by local experiments, taking proper account of the impending end of inflation.

While specific inflationary models abound in the literature, they can almost
entirely be brought under the umbrella of {\em chaotic inflation} \cite{LIN},
the situation where inflation is driven by a single scalar field $\phi$, the
{\em inflaton}, evolving classically in a potential which may be freely
chosen. Different inflation models correspond to different choices of this
potential, and we restrict ourselves to models which can be described in this
way. Inflation is formally defined as the condition that the (synchronous)
scale factor $a(t)$ of the universe is accelerating, $\ddot{a} > 0$.

Utilizing the scalar field as a time variable, the equations of motion in
Hamilton-Jacobi form are \cite{HJ}
\begin{eqnarray}
\label{e1}
(H'(\phi))^2 - \frac{12 \pi}{m_{Pl}^2} H^2(\phi) & = & -
	\frac{32\pi^2}{m_{Pl}^4} V(\phi) \, ,\\
\label{e2}
\dot{\phi} & = & - \frac{m_{Pl}^2}{4\pi} H'(\phi) \,,
\end{eqnarray}
where $H = \dot{a}/a$ is the Hubble parameter, $m_{Pl}$ is the Planck mass,
dots are time derivatives and primes $\phi$-derivatives. Throughout we take
$\dot{\phi} > 0$. As an alternative to $V(\phi)$, one can specify models by a
choice of $H(\phi)$, determining the corresponding $V(\phi)$ by
differentiation. This is greatly advantageous, as the gravitational wave
spectrum depends only on the behaviour of $H$ and so can be dealt with
directly. In general one cannot analytically find the $H(\phi)$ corresponding
to a given $V(\phi)$, but for our purposes one has no need to.

It is convenient, following \cite{EPSINF}, to go yet one step further and
define a quantity $\epsilon(\phi)$ as
\begin{equation}
\epsilon(\phi) = \frac{m_{Pl}^2}{4\pi} \left( \frac{H'(\phi)}{H(\phi)}
	\right)^2 \,.
\end{equation}
This function can be identified as measuring the validity of the so-called
slow-roll approximation as inflation progresses \cite{EPSINF}. Here we shall
not need recourse to this property. Let us note though that an arbitrary
inflation model can be specified by a choice of $\epsilon(\phi)$, with the
corresponding $H(\phi)$ obtained by quadrature through
\begin{equation}
H(\phi) = H_{{\rm end}} \exp \left( \int_{\phi}^{\phi_{{\rm end}}}
	\sqrt{\frac{4\pi\epsilon(\phi)}{m_{Pl}^2}} {\rm d} \phi \right) \,,
\end{equation}
where $H_{{\rm end}}$ is $H(\phi_{{\rm end}})$, the Hubble parameter at the
end of inflation. The condition for inflation, $\ddot{a} > 0$, is precisely
equivalent to $\epsilon(\phi) < 1$. Hence $\phi_{{\rm end}}$ is determined by
the condition $\epsilon(\phi_{{\rm end}}) = 1$.

Inflation generates both density perturbation (scalar) and gravitational wave
(tensor) spectra. The calculations giving rise to these spectra are well
known, and we shall simply quote the results. The key characteristic of
inflation is that a given {\em comoving} scale $k$ (that is, one stretched
with the expansion $a(t)$) grows in relation to the characteristic scale of
the expansion, the comoving Hubble length $H^{-1}/a$. The scales of interest
to us possess the following history. Sufficiently early in inflation a given
comoving scale is well inside the Hubble length. As inflation proceeds, the
scale is stretched beyond the Hubble length, at which point vacuum
fluctuations in the scalar and gravitational wave modes are `frozen-in',
their amplitude remaining fixed. For the gravitational wave modes, this effect
is commonly called parametric or superadiabatic amplification \cite{GRISH},
and can be interpreted in quantum terms as the creation of gravitons. Much
later after inflation, the modes re-enter the Hubble radius and give rise to
observable consequences. On a given scale, the amplitude of the modes is
determined by the physical conditions, and in particular the behaviour of the
Hubble parameter, at the time they left the Hubble radius during inflation.
The modes thus obey a `first out, last in' structure, so that modes on larger
scales were frozen-in earlier during inflation. The standard expressions for
the spectra of scalar and tensor modes are \cite{LL2}
\begin{eqnarray}
{\cal P}_S^{1/2}(k) & = & \frac{2}{m_{Pl}^2} \left.
	\frac{H^2(\phi)}{|H'(\phi)|} \right|_{aH=k} \,,\\
{\cal P}_T^{1/2}(k) & = & \left. \frac{4}{\sqrt{\pi}} \,
	\frac{H(\phi)}{m_{Pl}} \right|_{aH=k}\,,
\end{eqnarray}
where the evaluations are carried out when the comoving scale $k$ crossed
outside the Hubble radius.

The simplest quantity to compare with the experimental sensitivity of
gravitational wave detectors is the present-day contribution per octave of the
gravitational waves to the total energy density, given by \cite{OMG}
\begin{equation}
\Omega_g(k) \simeq \frac{2}{3\pi} \left( \frac{H}{m_{Pl}} \right)^2 \times 4
	\times 10^{-5} h^{-2} \,,
\end{equation}
where $H$ is evaluated as the scale $k$ crosses outside the Hubble radius. The
last factor accounts for redshifting during the matter-dominated era, and $h$
is the present-day Hubble parameter in units of $100 {\rm km s}^{-1} {\rm
Mpc}^{-1}$, assumed to lie between 0.4 and 1. For LIGO the peak sensitivity is
$\Omega_g \simeq 10^{-11}$ at $10$ Hz \cite{LIGO}, while the beam-in-space
sensitivity peaks at $\Omega_g \simeq 10^{-16}$ at $10^{-4}$ Hz \cite{BIS}.

One needs to connect the comoving scale to the value of $\phi$ during
inflation. This is carried out via the number of $e$-foldings $N(k)$ of
expansion between scale $k$ crossing the Hubble radius and the end of
inflation, defined by $N(k) = \ln \left( a(k)/a_{{\rm end}}\right)$. For our
purposes, the expression
\begin{equation}
\label{NK}
N(k) = 60 - \ln \frac{k}{a_0 H_0} \,,
\end{equation}
(`0' indicating present value) is easily adequate \cite{LL2}; the value 60
depends on the reheating properties after inflation, but our final results
have only a very weak dependence on it. The number of $e$-foldings between two
scalar field values, the latter taken as $\phi_{{\rm end}}$, is given exactly
by
\begin{equation}
\label{NPHI}
N(\phi,\phi_{{\rm end}}) = \sqrt{\frac{4\pi}{m_{Pl}^2}}
	\int_{\phi}^{\phi_{{\rm end}}} \frac{1}{\sqrt{\epsilon(\phi)}}
	\, {\rm d}\phi \,.
\end{equation}

The comoving scale equal to the present Hubble radius, $1/a_0 H_0 = 3000
h^{-1} \; {\rm Mpc} = 10^{28} h^{-1} \; {\rm cm}$, equalled the Hubble radius
around 60 $e$-foldings before the end of inflation. The beam-in-space
experiments are sensitive to much smaller scales, with peak sensitivity at
wavelengths around $10^{14}$ centimeters; this scale crossed the Hubble radius
33 $e$-foldings later. The corresponding figure for the LIGO experiment is 44.
In fact, for extreme model parameters the LIGO scales may never pass outside
the Hubble radius during inflation, and the gravitational wave production
there would be severely suppressed in accord with the adiabatic limit.

The COBE normalization fixes the integration constant which appears when one
goes from $\epsilon(\phi)$ to $H(\phi)$. It is easiest to specify that
integration constant as $H_{60}$, the value of $H$ when the presently
observable universe crossed outside the Hubble radius, rather than $H_{{\rm
end}}$. In cases where inflation is close to exponential, the standard COBE
normalization is ${\cal P}_S^{1/2}(k=a_0 H_0) = 4.3 \times 10^{-5}$. [The
quantity $\delta_H$ used in \cite{LL2,EPSINF} equals $(2/5) {\cal
P}_S^{1/2}$.] However, if $\epsilon_{60}$ or $\epsilon_{60}'$ are noticeably
different from unity then there are corrections which must be taken into
account, and this is done in accord with the treatment in \cite{EPSINF}.

An important guide is an approximate analytic measure of the relative
significance of the tensor and sca\-lar contributions to large angle microwave
background anisotropies. The relative contributions to the squared expectation
$\Sigma_l^2$ of the $l$-th multipole of a spherical harmonic decomposition can
be written as \cite{LL2}
\begin{equation}
\frac{\Sigma_l^2({\rm tensor})}{\Sigma_l^2({\rm scalar})} \simeq
	\frac{25}{2} \epsilon(\phi_l) \,,
\end{equation}
where $\phi_l$ indicates evaluation when the scales predominantly contributing
to the $l$-th multipole, $k = l a_0 H_0/2$, crossed the Hubble radius during
inflation. Consequently, tensor modes contribute significantly only if
$\epsilon(\phi_l)$ is at least a few hundredths.

In bounding $H_{60}$, one finds the initially promising result that the
largest values of $H_{60}$ come from choosing a large $\epsilon_{60}$. To
ensure consistency of the calculational method, mild restrictions on
deviations from slow-roll are made, as discussed in \cite{EPSINF}; with these
the largest COBE normalization of is $H_{60} = 2.9 \times 10^{-5} m_{Pl}$
occurring with $\epsilon_{60} = 0.25$, the maximum permitted. So the Hubble
parameter is largest at 60 $e$-foldings in cases where the gravitational wave
contribution is substantial. This somewhat surprising result occurs because of
the effect of the dramatic deviations of the spectra from the usual
scale-invariance on the normalization in this case \cite{EPSINF}. This
deviation has a detrimental effect on the anticipated local gravitational wave
signal.

As the Hubble parameter can only decrease as the expansion continues, this
sets a weak but absolute upper limit on $\Omega_g$ across all scales of
\begin{equation}
\Omega_g h^2 < 7 \times 10^{-15} \,,
\end{equation}
which is well below the LIGO sensitivity but comfortably within that of
the beam-in-space, especially for low $h$.

The next step is to implement some very weak assumptions regarding the nature
of inflation. If one has a large gravitational wave contribution to COBE,
one requires a large value of $H'$ and consequently has a rapidly decreasing
Hubble parameter. It would be possible for the Hubble parameter to
sharply level off as soon as scales contributing to COBE have crossed the
Hubble radius; such `designer' situations cannot be excluded on physical
grounds, but are physically unappealing.

Because of the way the problem has been set up, inflation must end within 60
$e$-foldings. This requires that $\epsilon(\phi)$ must grow to unity, which
will necessitate a reduction in $H$. The condition for 60 $e$-foldings is
\begin{equation}
\label{EFOLD}
\frac{60}{\sqrt{4\pi}} = \int_{\phi_{60}}^{\phi_{{\rm end}}}
	\frac{1}{\sqrt{\epsilon(\phi)}} \, \frac{{\rm d}\phi}{m_{Pl}} \,.
\end{equation}
The reduction in $H$ that occurs during the last 60 $e$-foldings subject to
this constraint is
\begin{equation}
\label{HRAT}
\frac{H_{{\rm end}}}{H_{60}} = \exp \left( - \sqrt{4\pi}
	\int_{\phi_{60}}^{\phi_{{\rm end}}} \sqrt{\epsilon(\phi)} \,
	\frac{{\rm d}\phi}{m_{Pl}} \right) \,,
\end{equation}
where $\epsilon(\phi_{{\rm end}}) = 1$ and $H_{60}$ is determined from the
COBE normalization for the given $\epsilon_{60}$ and $\epsilon_{60}'$. For
gravitational wave purposes, we are more interested in the reduction during
the first 33 of these $e$-foldings, which get us down to the beam-in-space
frequencies.

In order to gain meaningful results, one must impose some constraint on the
inflationary models, which is intended to represent some sense of `physical
reasonableness'. Such a notion is subject both to prejudice and to
exceptions, and so it is in one's interest to make as weak assumptions as one
feels able to. The choice made here is to assume that $\epsilon(\phi)$ does
not decrease during the last 60 $e$-foldings. In almost all models,
$\epsilon(\phi)$ must rise to unity, and were one to choose a functionally
complex $\epsilon(\phi)$ to violate this assumption then the underlying
potential would be complex, threatening the prejudice that the potential
belongs to a simple underlying theory. Further, almost all known models do fit
into this assumption, including chaotic inflation with polynomial potentials
\cite{LIN}, natural inflation \cite{NAT} and power-law inflation \cite{LM}. It
also includes models where inflation ends by an additional mechanism, and
which thus don't necessarily end at $\epsilon(\phi) = 1$, such as extended
inflation \cite{EI}. Those rare known examples which do not fit this
assumption, intermediate inflation \cite{BL} and some versions of models based
on two scales \cite{HYBRID}, can be shown case by case to produce low local
gravitational wave amplitudes unless extreme fine-tuning is introduced.

With this assumption, for a given $\epsilon_{60}$ one has
\begin{equation}
\frac{H_{27}}{H_{60}} \leq \exp \left( - 33 \epsilon_{60} \right) \,,
\end{equation}
where $H_{27}$ is the Hubble parameter appropriate for beam-in-space scales,
equality being achieved by retaining a constant $\epsilon(\phi)$ until 60
$e$-foldings pass and then increasing it rapidly to unity. The exponential
factor indicates that choosing a large $\epsilon_{60}$ to maximize the
gravitational wave contribution to COBE is not the best way to maximize the
contribution on shorter scales. With $H_{60}$ COBE normalized, the overall
maximum value is found for $\epsilon_{60} \simeq 1/66$, giving rise to
$H_{27}^{{\rm max}} \simeq 5.5 \times 10^{-6} m_{Pl}$ supplying a
`beam-in-space' signal of
\begin{equation}
\Omega_g^{{\rm max}} h^2 \simeq 2.4 \times 10^{-16} \,.
\end{equation}
This is within the anticipated sensitivity, and so one concludes that at least
within this rather general set of inflation models, it is possible to
have consistency with COBE and be able to see a signal with beam-in-space.

Now suppose however that the COBE signal is identified as having a sizeable
gravitational wave component. For example, the tentative results of
\cite{CBDES} indicate roughly equal scalar and tensor contributions, implying
$\epsilon_{60} \simeq 1/12$. In that case (numerically calculating the exact
COBE normalization \cite{EPSINF}) one finds $\Omega_g h^2 \leq 1.3 \times
10^{-17}$, out of reach of the beam-in-space. Only once the tensor
contribution is less than half the scalar one is the prediction comfortably
within the anticipated sensitivity.

For concreteness, let's end with the predictions of some specific COBE
normalized models. Chaotic inflation models with $V(\phi) \propto
\phi^{\alpha}$ ($\alpha$ an even integer) give very small gravitational wave
contributions to COBE. Here the standard normalization and the slow-roll
approximation are adequate and yield for the beam-in-space
\begin{eqnarray}
\Omega_g h^2 & = & \frac{4.8 \times 10^{-14} \alpha}{240+\alpha}
	\left( \frac{108+\alpha}{240+\alpha} \right)^{\alpha/2} \,,\\
 & = & \left\{  \begin{array}{l}
		1.8 \times 10^{-16} \quad {\rm for}\;\; \alpha = 2 \,,\\
 		1.7 \times 10^{-16} \quad {\rm for}\;\; \alpha = 4 \,.
                \end{array} \right.
\end{eqnarray}
This is comfortably detectable and indeed not far from the optimal theoretical
value. Power-law models $a(t) \propto t^p$ with high powers give results close
to the optimal with negligible gravitational effect on COBE. But low powers,
which give a large effect on COBE, invariably give smaller beam-in-space
predictions. One can solve the Hamilton-Jacobi equations exactly for $H(\phi)$
and numerically calculate the COBE normalization to find
\begin{equation}
\Omega_g h^2  = \left\{ \begin{array}{l}
	2.3 \times 10^{-16} \quad {\rm for}\;\; p \simeq 100 \,,\\
	4.3 \times 10^{-18} \quad {\rm for}\;\; p \simeq 10 \,.
	\end{array} \right.
\end{equation}

In conclusion, both abstract and particular models indicate a reasonable
expectation that beam-in-space experiments can see the stochastic
gravitational wave background, particularly if the present Hubble constant
proves to be low. Predictions are however generically below $\Omega_g =
10^{-15}$ even for $h = 0.5$, indicating that the full anticipated sensitivity
will be required. Counter-intuitively, the larger the gravitational wave
signal in the microwave background, the smaller the expected signal for the
beam-in-space experiment.

The author is supported by the SERC, and acknowledges the use of the Starlink
computer system at the University of Sussex. I am grateful to Jim Hough for a
conversation prompting me to look at this question, and to John Barrow,
Alfredo Henriques, Gordon Moorhouse and David Wands for discussions.
\frenchspacing

\end{document}